# Comments on "Maximally permissive supervisor synthesis based on a new constraint transformation method"[Automatica 48 (2012) 1097-1101]

Shouguang Wang[a], Jing Yang[a] and Mengchu Zhou[b]

[a]*Zhejiang Gongshang University, Hangzhou, China*
*{wangshouguang,jyxd}@mail.zjgsu.edu.cn*

[b]*New Jersey Institute of Technology, Newark, NJ 07102, USA*
*zhou@njit.edu*

**Abstract**

Luo *et al*. proposed a new method to design the maximally permissive and efficient supervisor for enforcing linear constraints on an ordinary Petri net with uncontrollable transitions. In order to develop this method, Theorem 3 is given. It is clamed that "a linear constraint can be equivalently transformed at an uncontrollable transition into a disjunction of weakly admissible ones." However, this result is erroneous. In this correspondence paper, a counterexample contradicting it is presented.

## 1. Introduction

Recently, Luo *et al*. (2012) have presented a new method to perform the constraint transformation from a linear constraint into a set of weakly admissible ones for an ordinary Petri net (PN) with uncontrollable transitions. The research on how to implement linear constraints of PNs which can describe many discrete event system control problems is a valuable topic to deal with the control issues in discrete event systems due to their broad range of applications in manufacturing, transportation, process automation, and so on. Some studies have reported suitable solutions if each transition is controllable. Others have focused on the issue when there exist uncontrollable transitions. However, the linear constraint transformation on an ordinary PN with uncontrollable transitions is still a difficult and open topic, which needs to be further studied.

Luo *et al*. (2012) provide us a new method on this open issue and solved some linear constraint transformation problems given in their work. To develop the proposed method, three theorems are given. The proposed theorems are proven and suitable for the examples given in (Luo *et al*., 2012).

After our thorough study of their results, we have identified an error in one of their theorems, which states that a linear constraint is equivalent to the disjunction of new linear constraints transformed. This result is not applicable to all ordinary PNs, which is demonstrated by using a counterexample in the next section.

## 2. Main Results

In this section, Theorem 3 in Luo *et al*. (2012) is briefly reviewed. Then, a counterexample is presented. Here, the same notations are adopted as those in Luo *et al*. (2012).

Let $\mathcal{N} = (\mathcal{P}, \mathcal{T}, \mathcal{F}, \mathcal{W})$ denote an ordinary PN, where $\mathcal{P}$ (places) and $\mathcal{T}$ (transitions) are disjoint sets representing the graph vertices, $\mathcal{F} \subseteq (\mathcal{P} \times \mathcal{T}) \cup (\mathcal{T} \times \mathcal{P})$ is a set of direct arcs connecting places and transitions, and $\mathcal{W}: \mathcal{F} \to \mathbb{Z}^+$ is a mapping that assigns a positive integer weight to each arc in $\mathcal{F}$. $\mathcal{N}$ is ordinary if $\mathcal{W}(f) = 1$ for all $f \in \mathcal{F}$.

($w$, $k$) stands for a linear constraint to be implemented on a PN, where $w$ is a function from $\mathcal{P}$ to $\mathbb{Z}$ and $k$ is an integer. $w(p)$ is called the weight of $p$. It is assumed that the weight of each place is not negative. A linear constraint $(w, k)$ requires that the marking $m$ of a PN satisfy $w \cdot m \leq k$.

Let $\mathcal{T}_c(\mathcal{T}_{uc})$ denote the set of controllable (uncontrollable) transitions. $^\bullet t$ and $t^\bullet$ represent the set of input and output places of $t$, respectively. $R(m)$ denotes the set of markings that are reachable from $m$. The legal-marking set is denoted as $\mathcal{L}_{\omega,k} := \{m \in R(m_0) | w \cdot m \leq k\}$. The admissible-marking set is denoted as

$$\mathcal{A}_{w,k} := \{m \in R(m_0) | R_{\mathcal{T}_{uc}}(m) \subseteq \mathcal{L}_{w,k}\}$$

where $R_{\mathcal{T}_{uc}}(m)$ is the set of markings that are reachable by firing only uncontrollable transitions from $m$. Let $W = \{(w_1, w_k), \ldots, (w_n, k_n)\}$, $n \in \mathbb{Z}^+$, denote a set of linear constraints. The disjunction of the constraints in $W$ is denoted as $\vee(W)$, that is, $\vee_{(w,k)\in W} w \cdot m \leq k$. Its legal-marking set is $\mathcal{L}_{\vee(W)} := \bigcup_{(w,k)\in W} \mathcal{L}_{w,k}$, and its admissible-marking set is $\mathcal{A}_{\vee(W)} := \bigcup_{(w,k)\in W} \mathcal{A}_{w,k}$. The following definitions, lemma and theorem are cited from (Luo *et al.*, 2012).

**Definition 1.** Given a PN with linear constraint $(w, k)$, the weight of transitions is defined as a row vector, $\varpi = w \cdot [\mathcal{N}]$.

From Definition 1, $\varpi(t) = \sum_{p \in t^\bullet} w(p) - \sum_{p \in {}^\bullet t} w(p)$. $(w, k)$ is admissible if $\forall t \in \mathcal{T}_{uc}, \varpi(t) \leq 0$ (Moody & Antsaklis, 2002).

**Definition 2.** $(w, k)$ is a weakly admissible linear constraint if $\forall p \in \mathcal{P}$, $w(p) \geq 0$, and $\forall t \in \mathcal{T}_{uc}, \varpi(t) > 0 \Rightarrow (\exists p \in {}^\bullet t, w(p) > k)$.

**Definition 3.** The uncontrollable transition gain transformation function is $\rho: \mathcal{LW} \times \mathcal{T}_{uc} \times \mathcal{P} \to \mathcal{LW}$, where $\mathcal{LW}$ is the linear-constraint set. It is defined as

$\forall (w,k) \in \mathcal{LW}$, $\forall t \in \mathcal{T}_{uc}$, $\forall p \in \mathcal{P}$, $(w',k') = \rho((w,k),t,p)$, such that

$$\begin{cases} k' = k \\ \forall p' \in \mathcal{P}, \quad w'(p') = \begin{cases} w(p'), & \text{if } p \notin {}^\bullet t \vee p' \neq p; \\ w(p') + w, & \text{Otherwise.} \end{cases} \end{cases}$$
(1)

**Definition 4.** Given an uncontrollable transition $t$ whose weight is positive, that is, $\varpi(t) > 0$, then $t$'s complement weight set (CWS) is defined as

$$\varrho((w,k),t) = \bigcup_{p \in {}^\bullet t} \{\rho((w,k),t,p)\},$$
(2)

where $\rho$ is the uncontrollable transition gain transformation function from Definition 3.

**Definition 5.** If any marking of $(\mathcal{N}, m_0)$ is not admissible for a linear constraint $(w, k)$, then $(w, k)$ is called the zero constraint denoted as $\mathbf{0}$.

The disjunction of $(w, k)$ and the zero constraint is equivalent to $(w, k)$, that is, $(w, k) \vee \mathbf{0} \equiv (w, k)$.

**Lemma 3.** If $w \cdot m_0 > k$, then $(w, k) \equiv \mathbf{0}$.

**Theorem 3.** If $(\mathcal{N}, m_0)$ and $(w, k)$ are the inputs of Algorithm 1 and $W$ is the output, then
(a) Each linear constraint in $W$ is weakly admissible constraint for $(\mathcal{N}, m_0)$;
(b) The original constraint $(w, k)$ is equivalent to the disjunction of the linear constraints in $W$, i.e. $(w, k) \equiv \vee(W)$.

With careful examination of the applications of Theorem 3, it is found that this theorem is incorrect. To show it, the following counterexample is given in Fig. 1.

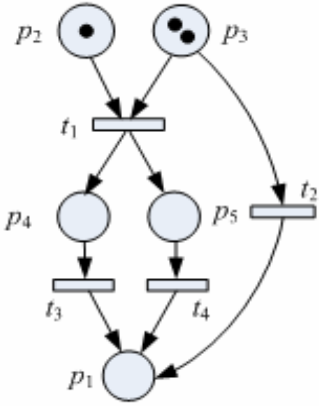

Fig. 1. A Petri net used as a counterexample.

For the convenient representation of the counterexample, the definition of $\varrho((w,k),t)$ from Definition 4 is expanded such that $\varrho((w,k),t)$ is defined under the case $\varpi(t) \leq 0$.

**Definition 6:** Given an uncontrollable transition $t$, a linear constraint $(w,k)$ and linear constraint set $W$, $\varrho((w,k),t)$ is redefined as

$$\varrho((w,k),t) = \begin{cases} (w,k), & \varpi(t) \leq 0 \\ \bigcup_{p \in {}^\bullet t}\{\rho((w,k),t,p)\}, & \varpi(t) > 0 \end{cases} \quad (3)$$

$\varrho(W,t)$ is defined as

$$W_t = \varrho(W,t) = \bigcup_{(w,k) \in W} \varrho((w,k),t) \quad (4)$$

**Counterexample**: Consider the linear constraint $(w, 3)$ on $(N, m_0)$ in Fig. 1, where $t_1, t_2, t_3$ and $t_4$ are uncontrollable, $w = (1,0,0,1,1)$, and the initial marking is $m_0 = (0, 1, 2, 0, 0)^T$.

It is noted that there are two different ways to implement the linear constraint transformation for this counterexample: Case 1, implementing the transformation starting with $t_1$ first and then $t_2$; and Case 2, implementing the transformation starting with $t_2$ first and then $t_1$.

1) **Case 1**: Transforming the original constraint $(w, 3)$ by $t_1$ first and then by $t_2$.

(a) $W_{t_1} = \varrho((w,3),t_1)$
$= \{\rho((w,3),t_1,p_2), \rho((w,3),t_1,p_3)\}$
$= \{(w_1,3),(w_2,3)\}$

where $w_1 = (1,2,0,1,1)$ and $w_2 = (1,0,2,1,1)$.

(b) $W_{t_1 t_2} = \varrho(W_{t_1}, t_2)$
$= \{\varrho((w_1,3),t_2), \varrho((w_2,3),t_2)\}$
$= \{(w_3,3),(w_2,3)\}$

where $w_3 = (1,2,1,1,1)$.

2) **Case 2**: Transforming the original constraint $(w, 3)$ by $t_2$ first and then by $t_1$.

(a) $W_{t_2} = \varrho((w,3),t_2)$
$= \rho((w,3),t_2,p_3)$
$= \{(w_4, 3)\}$

where $w_4 = (1,0,1,1,1)$.

(b) $W_{t_2 t_1} = \varrho(W_{t_2}, t_1)$
$= \{\rho((w_4,3),t_1,p_3), \rho((w_4,3),t_1,p_2)\}$
$= \{(w_5,3),(w_6,3)\}$

where $w_5 = (1,0,2,1,1)$ and $w_6 = (1,1,1,1,1)$.

If Theorem 3 were correct, the following facts would be deduced from it:

**For case 1**: $(w,k) \equiv \vee(W_{t_1 t_2})$ that is, $\mathcal{A}_{w,k} = \mathcal{A}_{\vee(W_{t_1 t_2})}$.

**For case 2**: $(w,k) \equiv \vee(W_{t_2 t_1})$ that is, $\mathcal{A}_{w,k} = \mathcal{A}_{\vee(W_{t_2 t_1})}$.

Therefore,

$$\mathcal{A}_{\vee(W_{t_1 t_2})} = \mathcal{A}_{\vee(W_{t_2 t_1})} \quad (5)$$

Next, let us check whether (5) is true or not.

**For Case 1:**

Because $w_2 \cdot m_0 = 4 > 3$ and $w_3 \cdot m_0 = 4 > 3$, $(w_2,3) \equiv 0$ and $(w_3,3) \equiv 0$ from Lemma 3. Therefore,

$$\mathcal{A}_{\vee(W_{t_1 t_2})} = \varnothing \quad (6).$$

**For Case 2:**

Because $w_5 \cdot m_0 = 4 > 3$, $(w_5,3) \equiv 0$ from Lemma 3, Therefore,

$$\mathcal{A}_{\vee(W_{t_2 t_1})} = \mathcal{A}_{w_6,3} \neq \varnothing \quad (7).$$

Note that $m_0 \in \mathcal{A}_{w_6,3}$. From (6) and (7), it is clear that $\mathcal{A}_{\vee(w_{t_1 t_2})} \neq \mathcal{A}_{\vee(w_{t_2 t_1})}$, which means that (5) is not true for this example. Since (5) is derived from Theorem 3 and it is not true, it can be concluded that Theorem 3 is incorrect and not suitable for the presented example.

## 3. Conclusion

In this note, it has been shown that Theorem 3 in (Luo *et al*., 2012) is incorrect from the given counterexample. Theorem 3 claims that the original linear constraint can be equivalently transformed at an uncontrollable transition into a disjunction of weekly admissible ones. However, it fails to consider that different transformation sequences of uncontrollable transitions might result in non-equivalent outputs, as shown by the given counterexample.

## References

Luo, J., Shao, H., Nonami, K., & Jin, F. (2012) Maximally permissive supervisor synthesis based on a new constraint transformation method. *Automatica*, 48, 1097-1101

Moody, J. O., & Antsaklis, P. J., (2000). Petri Net supervisiors for DES with uncontrollable and unobservable transitions. *IEEE Trans. Automat. Control*. 45(3), 462-476.